\documentclass[twocolumn,superscriptaddress,longbibliography,aps,prb,preprintnumbers,nofootinbib,showkeys]{revtex4-2}
\usepackage[utf8]{inputenc}
\usepackage{graphicx}
\usepackage{bm}
\usepackage{amsmath}
\usepackage{amssymb}
\usepackage{hyperref}
\usepackage{subfigure}
\usepackage{float}
\usepackage{multirow}
\usepackage{booktabs}
\usepackage{varioref}
\usepackage{xr-hyper}
\usepackage[dvipsnames]{xcolor}
\usepackage{nicefrac}
\usepackage{xfrac}
\usepackage{hyperref}
\hypersetup{colorlinks,linkcolor=blue,urlcolor=blue,citecolor=blue}
\usepackage{ulem}
\usepackage{siunitx}
\usepackage{graphicx}
\usepackage{dcolumn}
\usepackage{braket}
\usepackage{wasysym}
\usepackage{textcomp}
\usepackage{pdfpages}

\newcommand{\LCO}{{La}$_{2}${CuO}$_4$}
\newcommand{\LNO}{{La}$_{2}${NiO}$_4$}

\newcommand{\dz}{$d_{z^2}$}
\newcommand{\dx}{$d_{x^2-y^2}$}

\makeatletter
\AtBeginDocument{\let\LS@rot\@undefined}
\makeatother

\begin{document}

\title{Strain-Tuned Incompatible Magnetic Exchange-Interaction in La$_2$NiO$_4$}

\author{Izabela~Bia\l{}o}
\email{izabela.bialo@uzh.ch}
\affiliation{Physik-Institut, Universit\"{a}t Z\"{u}rich, Winterthurerstrasse 
190, CH-8057 Z\"{u}rich, Switzerland}
\affiliation{AGH University of Krakow, Faculty of Physics and Applied Computer Science, 30-059 Krakow, Poland}

\author{Leonardo~Martinelli}
\affiliation{Physik-Institut, Universit\"{a}t Z\"{u}rich, Winterthurerstrasse 190, CH-8057 Z\"{u}rich, Switzerland}

\author{Gabriele~De~Luca}
\affiliation{Institut de Ciència de Materials de Barcelona (ICMAB-CSIC), 08193 Bellaterra (Barcelona), Spain}

\author{Paul~Worm}
\affiliation{Institute of Solid State Physics, Vienna University of Technology, A-1040 Vienna, Austria}

\author{Annabella~Drewanowski}
\affiliation{Physik-Institut, Universit\"{a}t Z\"{u}rich, Winterthurerstrasse 190, CH-8057 Z\"{u}rich, Switzerland}

\author{Simon~J\"{o}hr}
\affiliation{Physik-Institut, Universit\"{a}t Z\"{u}rich, Winterthurerstrasse 190, CH-8057 Z\"{u}rich, Switzerland}

\author{Jaewon~Choi}
\affiliation{Diamond Light Source, Harwell Campus, Didcot, Oxfordshire OX11 0DE, United Kingdom}

\author{Mirian~Garcia-Fernandez}
\affiliation{Diamond Light Source, Harwell Campus, Didcot, Oxfordshire OX11 0DE, United Kingdom}

\author{Stefano~Agrestini}
\affiliation{Diamond Light Source, Harwell Campus, Didcot, Oxfordshire OX11 0DE, United Kingdom}

\author{Ke-Jin~Zhou}
\affiliation{Diamond Light Source, Harwell Campus, Didcot, Oxfordshire OX11 0DE, United Kingdom}

\author{Kurt~Kummer}
\affiliation{ESRF, The European Synchrotron, 71 Avenue des Martyrs, CS40220, 38043 Grenoble Cedex 9, France}

\author{Nicholas~B.~Brookes}
\affiliation{ESRF, The European Synchrotron, 71 Avenue des Martyrs, CS40220, 38043 Grenoble Cedex 9, France}

\author{Luo~Guo}
\affiliation{Department of Materials Science and Engineering, University of Wisconsin-Madison, Madison, 53706, Wisconsin, USA}

\author{Anthony~Edgeton}
\affiliation{Department of Materials Science and Engineering, University of Wisconsin-Madison, Madison, 53706, Wisconsin, USA}

\author{Chang~B.~Eom}
\affiliation{Department of Materials Science and Engineering, University of Wisconsin-Madison, Madison, 53706, Wisconsin, USA}

\author{Jan~M.~Tomczak}
\affiliation{Department of Physics, King’s College London, Strand, London WC2R 2LS, United Kingdom}
\affiliation{Institute of Solid State Physics, Vienna University of Technology, A-1040 Vienna, Austria}

\author{Karsten~Held}
\affiliation{Institute of Solid State Physics, Vienna University of Technology, A-1040 Vienna, Austria}

\author{Marta~Gibert}
\affiliation{Institute of Solid State Physics, Vienna University of Technology, A-1040 Vienna, Austria}

\author{Qisi~Wang}
\email{qwang@cuhk.edu.hk}
\affiliation{Department of Physics, The Chinese University of Hong Kong, Shatin, Hong Kong, China}
\affiliation{Physik-Institut, Universit\"{a}t Z\"{u}rich, Winterthurerstrasse 190, CH-8057 Z\"{u}rich, Switzerland}

\author{Johan~Chang}
\affiliation{Physik-Institut, Universit\"{a}t Z\"{u}rich, Winterthurerstrasse 190, CH-8057 Z\"{u}rich, Switzerland}


\maketitle

\noindent\textbf{Abstract} \\
\textbf{Magnetic frustration is a route for novel ground states, including spin liquids and spin ices. Such frustration can be introduced through either lattice geometry or incompatible exchange interactions. Here, we find that epitaxial strain is an effective tool for tuning antiferromagnetic exchange interactions in a square-lattice system. By studying the magnon excitations in La$_2$NiO$_4$ films using resonant inelastic x-ray scattering, we show that the magnon displays substantial dispersion along the antiferromagnetic zone boundary, at energies that depend on the lattice of the film's substrate. Using first principles simulations and an effective spin model, we demonstrate that the antiferromagnetic next-nearest neighbour coupling is a consequence of the two-orbital nature of La$_2$NiO$_4$. Altogether, we illustrate that compressive epitaxial strain enhances this coupling and, as a result, increases the level of incompatibility between exchange interactions within a model square-lattice system.}
\\

\noindent\textbf{Introduction} \\
The square-lattice Heisenberg model is the subject of intense numerical and experimental investigations. In spin-1/2 systems---such as cuprates~\cite{LeeRMP2006} and copper deuteroformate tetradeurate~(CFTD)~\cite{ChristensenPNAS2007}---higher-order exchange interactions are inferred from observations of magnon dispersions along the magnetic zone boundary~\cite{HeadingsPRL2010,ColdeaPRL2001}. While a detailed magnon characterization is useful to understand quantum-fluctuation effects~\cite{Wang23}, exchange incompatibility is typically avoided in these systems. Indeed, the antiferromagnetic (AF) nearest-neighbor (NN) exchange interaction ($J_1>0$) and the ferromagnetic next-nearest-neighbour (NNN) interaction ($J_2<0$) in these systems stabilize the classical AF N\'eel order. Instead, magnetic exchange incompatibility requires both $J_1>0$ and $J_2>0$. This regime of the $J_1$-$J_2$ model is the subject of extensive computational investigations for both spin $S=1/2$~\cite{ChooPRB2019,CapriottiPRL2000,capriotti2001resonating, ZhangPRL2003,GongPRL2014} and $S=1$~\cite{SushkovPRB2001,JiangPRB2009} systems. In a narrow range near $J_2/J_1\sim \nicefrac{1}{2}$, magnetic frustration is found to dominate, and exotic quantum phases such as the spin-liquid state~\cite{Anderson1987} are predicted. Several calculations show that the Néel order is destroyed there and the ground state has a valence-bond character ~\cite{Anderson1987, dagotto1989phase, schulz_ziman_poilblanc_1996}, although its exact nature is still the subject of debate~\cite{beach2009master, capriotti2001resonating}. However, only very few square-lattice systems exhibit substantial magnetic frustration~\cite{WangNC2016,GuPRB2022}, and even fewer display tunable magnetic interactions~\cite{MustonenNC2018}. As a result, approaching the interesting parameter regime in real materials remains an ongoing issue.

\begin{figure*}
\centering
\includegraphics[scale=1]{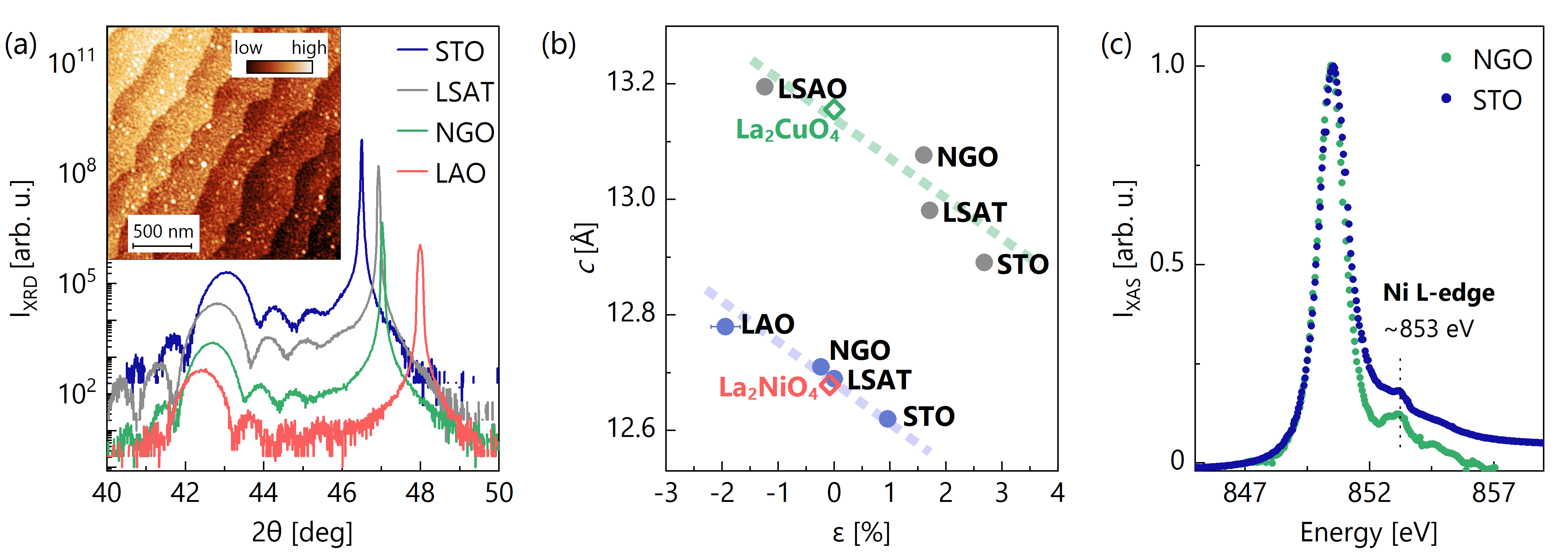}
\caption{\textbf{Characterization measurements on thin films of La$_2$NiO$_4$.} \textbf{(a)} X-ray diffraction (at $300$~K) probing the $(0,0,\ell)$ direction of $12$~nm thin films of La$_2$NiO$_4$ on substrates as indicated. The inset represents the atomic force microscopy image showing the step-like morphology of the films (here for the NGO substrate). The color scale corresponds to film thickness. \textbf{(b)} \textit{c}-axis lattice parameter versus in-plane epitaxial strain (at $300$~K) calculated for \LNO (purple dots) and La$_2$CuO$_4$~\cite{IvashkoNC2019} (gray dots) films grown on different substrates as a relative change of in-plane parameters in reference to  bulk (diamonds)  with $a = 3.868$~\AA\ and $c = 12.679$~\AA\ for an isomorphic La$_2$NiO$_4$ structure~\cite{Goodenough1982} and $a = 3.803$~\AA\ and $c = 13.156$~\AA\ for La$_2$CuO$_4$~\cite{Radaelli1994}. The errorbar correspond to the spread of in-plane parameters measured by x-ray diffraction. \textbf{(c)} X-ray absorption spectra around the Ni $L$-edge. The dominant peak corresponds to the La $M$-edge. (b,c) Dashed lines are guides to the eye.}
\label{Fig1}
\end{figure*}

In this article, we provide a high resolution resonant inelastic x-ray scattering (RIXS) study of magnetic excitations in epitaxial thin films of the canonical $S=1$ system La$_2$NiO$_4$, grown on different substrates. We discover a marked, upward dispersion along the AF zone-boundary $(\nicefrac{1}{2},0)\rightarrow(\nicefrac{1}{4}, \nicefrac{1}{4})$, which reveals the presence of AF NNN interactions that partly frustrate the NN ones. By employing \textit{ab initio} calculations, we demonstrate that these results can only be explained by including the multi-orbital nature of $3d^8$-Ni systems. Furthermore, we observe a correlation between the relative strength of the magnetic interactions and the strain applied onto the films. Our results demonstrate that 214-type nickelates are a promising class of materials for the study of the AF square-lattice Heisenberg model. Moreover, the use of thin films provides a clear route to tune the magnetic frustration and explore so far inaccessible regions of the magnetic phase diagram.\\

\noindent\textbf{Results} \\
Our thin films of La$_2$NiO$_4$ (LNO) on SrTiO$_3$ (STO), LaAlO$_3$ (LAO), (LaAlO$_3$)$_{0.3}$(Sr$_2$TaAlO$_6$)$_{0.7}$ (LSAT) and NdGaO$_3$ (NGO) substrates are characterized by atomic force microscopy, x-ray diffraction, and x-ray absorption spectroscopy -- see Fig.~\ref{Fig1}. The atomic force microscopy images display a steplike morphology indicating an excellent layer-by-layer growth. Diffraction patterns probing the $(0,0,\ell)$ reciprocal direction demonstrate good single crystallinity and allow us to extract the $c$-axis lattice parameters of the films. The epitaxial strain applied by the substrates is supported by the film $c$-axis and in-plane lattice-parameter dependence (Fig.~\ref{Fig1}b). X-ray absorption spectra recorded on the LNO/STO and LNO/NGO, shown in Fig.~\ref{Fig1}c, are consistent with observations on related nickelates~\cite{NagPRL2020,GhiringhelliNJP2005,LinPRL2021,KuiperPRB1998}. The Ni $L$-edge features on the tail of the La $M$-edge.

\begin{figure*}
\includegraphics[scale=1]{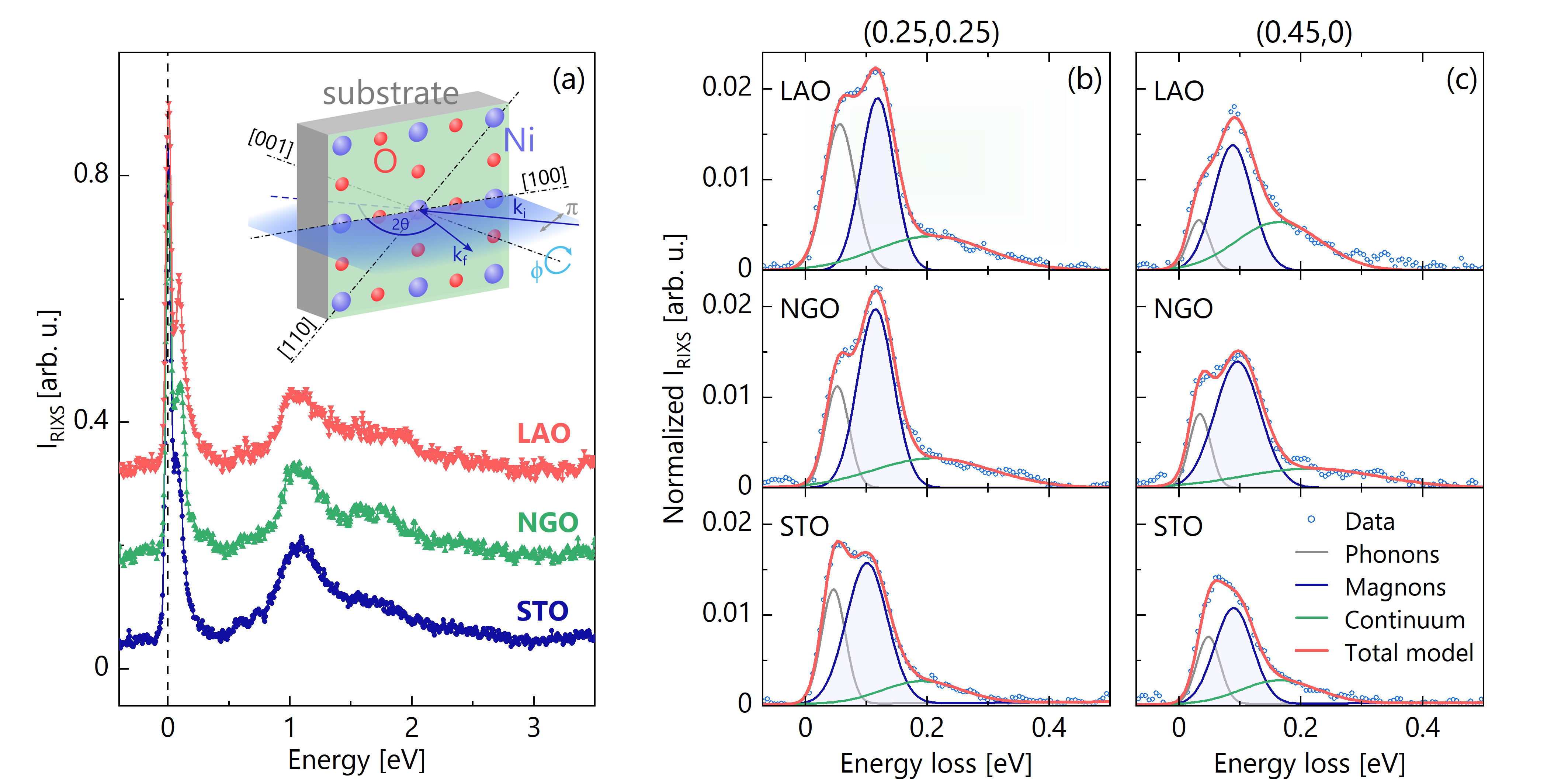}
\caption{\textbf{Resonant inelastic x-ray scattering spectra of \LNO.} \textbf{(a)} Raw spectra recorded in \LNO films with substrates as indicated. \textbf{(a,inset)} Schematics of the photon-in-photon-out resonant inelastic x-ray scattering (RIXS) geometry with horizontally polarized light~($\pi$) and azimuthal sample rotation angle $\phi$. \textbf{(b,c)} Low-energy part of the RIXS spectra with momentum transfer and film substrates indicated. The solid red line indicates a three-component fit with phonon, magnon (shaded), and multi-magnon (continuum) contributions. The elastic scattering channel is subtracted in (b,c).}
\label{Fig2}
\end{figure*}

The RIXS spectra of \LNO films were measured at the Ni $L_3$ edge ($853$~eV). These spectra exhibit key RIXS excitations, including high-energy $dd$-excitations (at approximately $0.5$-$3$~eV), an elastic scattering contribution at $0$~eV, as well as phonon and magnon excitations in between. The $dd$-excitations have a multi-peak structure, qualitatively similar to other $3d^8$ systems, such as NiO~\cite{ghiringhelli2009observation, GhiringhelliNJP2005, NagPRL2020, LinPRL2021}. As shown in Fig.~\ref{Fig2}a, the relative intensities of the peaks are different in the three samples, due to the different crystal-fields acting on the Ni atoms (see Fig.~S1, Supplementary Note~1). However, all our La$_2$NiO$_4$ films display the most intense $dd$-excitation around 1.1~eV and a second less intense excitation just below 1.6~eV. This is consistent with what is reported in bulk La$_2$NiO$_4$~\cite{FabbrisPRL2017} (see Fig.~S2, Supplementary Note~1). The subtraction of the elastic peak clearly highlights the presence of multiple low-energy features; see Figs.~\ref{Fig2}b,c. 

\begin{figure*}
\includegraphics[width={\linewidth}]{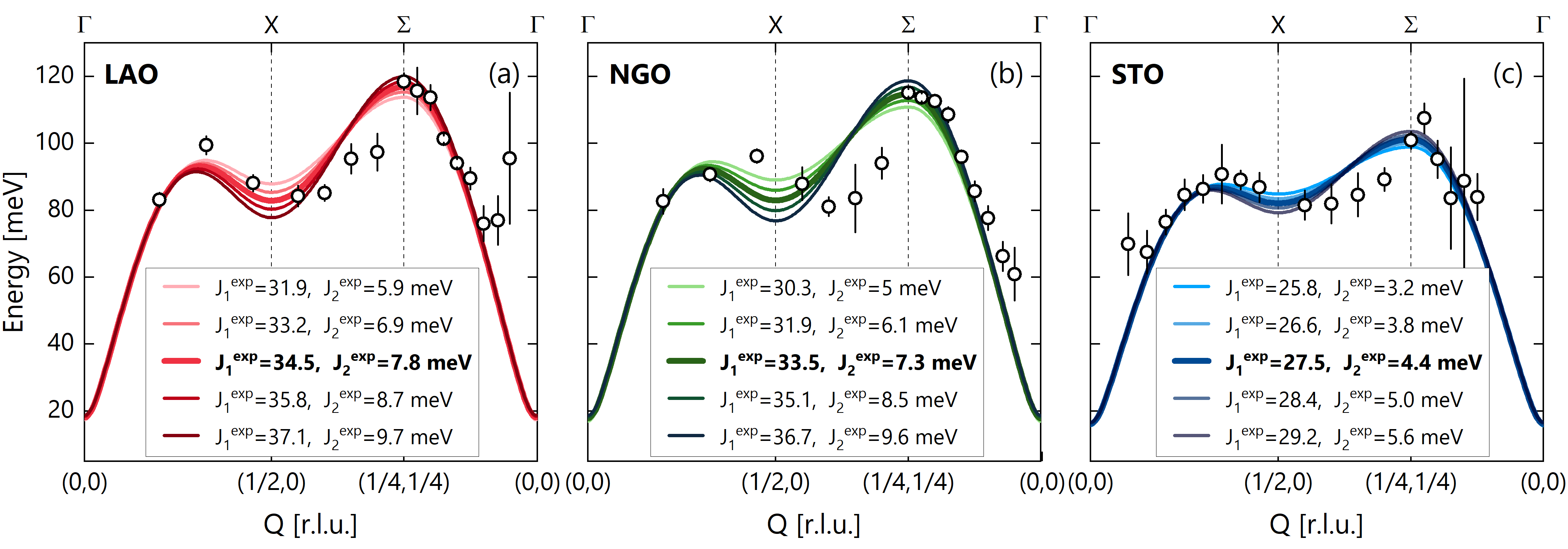}
\caption{\textbf{Magnon dispersion of \LNO\ films.} \textbf{(a-c)} Magnon excitation energies (open dots) along high symmetry directions for \LNO\ on substrates as indicated. Solid lines represent the same spin wave model evaluated for different exchange parameters within the confidence intervals of the fitted parameters. The curves corresponding to the best-fit values (marked in bold in the legend) are reported as thicker lines. The middle segment, $X\rightarrow\Sigma$, is part of the antiferromagnetic zone-boundary. The error bars are determined from the fitting uncertainty.
}
\label{fig_SpinWave}
\end{figure*}

To extract the dispersion of magnetic excitations, we assumed a two-mode model with the addition of a high-energy continuum ``background" (Fig.~\ref{Fig2}b,c). Each of these components is represented by a Gaussian profile. This provides an effective fitting model of excitations for all measured film systems and momenta. Our interpretation of the proposed model is based on the hypothesis that the lower-energy mode ($\sim$40~meV) stems from an optical phonon, while the higher-energy mode (strongly dispersing between $60$ and $120$~meV) is a magnon. This assignment is supported by previous neutron scattering measurements that identified the phonon part via an out-of-plane oxygen buckling mode~\cite{pint1989lattice, pist1988phonon}. The interpretation of the higher-energy mode as a magnon is consistent with earlier RIXS~\cite{FabbrisPRL2017} and neutron studies~\cite{nakajima1993spin} of bulk La$_2$NiO$_4$. The resulting magnon dispersions are shown in Fig.~\ref{fig_SpinWave}.

Due to lower energy resolution, the previous RIXS study~\cite{FabbrisPRL2017} did not resolve any phonon excitations. The unresolved phonon excitation implied that the phonon and magnon spectral weights were merged. This, in turn influences the extraction of the magnon dispersion. Having access, in this work, to a higher energy resolution, we can distinguish between the nearly momentum-independent phonon mode and the dispersive magnon branch along the three measured high-symmetry directions. In all the film systems explored, the magnon energy reaches its maximum at the AF zone boundary, at the $\Sigma$ point $(\nicefrac{1}{4},\nicefrac{1}{4})$, referred to as $E_{\Sigma}$, while it displays a local minimum at the $X$ point $(\nicefrac{1}{2},0)$, referred to as $E_{X}$. This evidently anisotropic shape of magnon dispersion was not reported in earlier studies~\cite{FabbrisPRL2017,nakajima1993spin}, except for a recent inelastic neutron scattering experiment~\cite{Petsch2023}. Furthermore, the energy $E_{\Sigma}$ is different for all three substrates. In particular, it increases as a function of compressive strain, with an enhancement of $18\pm4\:$meV ($\sim20\%$) from LNO/STO to LNO/LAO. 
\\

\noindent\textbf{Discussion} \\
By resolving both the phonon and magnon modes, we find that all samples exhibit a substantial dispersion of magnetic excitations along the AF zone boundary. This directly implies the presence of higher-order effective magnetic exchange interactions. In La$_2$CuO$_4$ and related Mott insulating cuprates, the zone boundary dispersion has been interpreted in terms of a positive ring-exchange interaction that emerges naturally from a single-orbital Hubbard model~\cite{peng2017influence,IvashkoPRB2017,IvashkoNC2019}. There is, however, an important difference between the zone-boundary dispersion of La$_2$CuO$_4$ and La$_2$NiO$_4$: in contrast to La$_2$CuO$_4$, the zone boundary dispersion of La$_2$NiO$_4$ has its maximum at the AF zone boundary $\Sigma$ point rather than at the $X$ point. As such, the magnon dispersion of La$_2$NiO$_4$ is (as could be expected) inconsistent with a single-band Hubbard model in the strong coupling limit (where the projection onto a Heisenberg spin Hamiltonian is viable).

\begin{figure}
\includegraphics[width={\linewidth}]{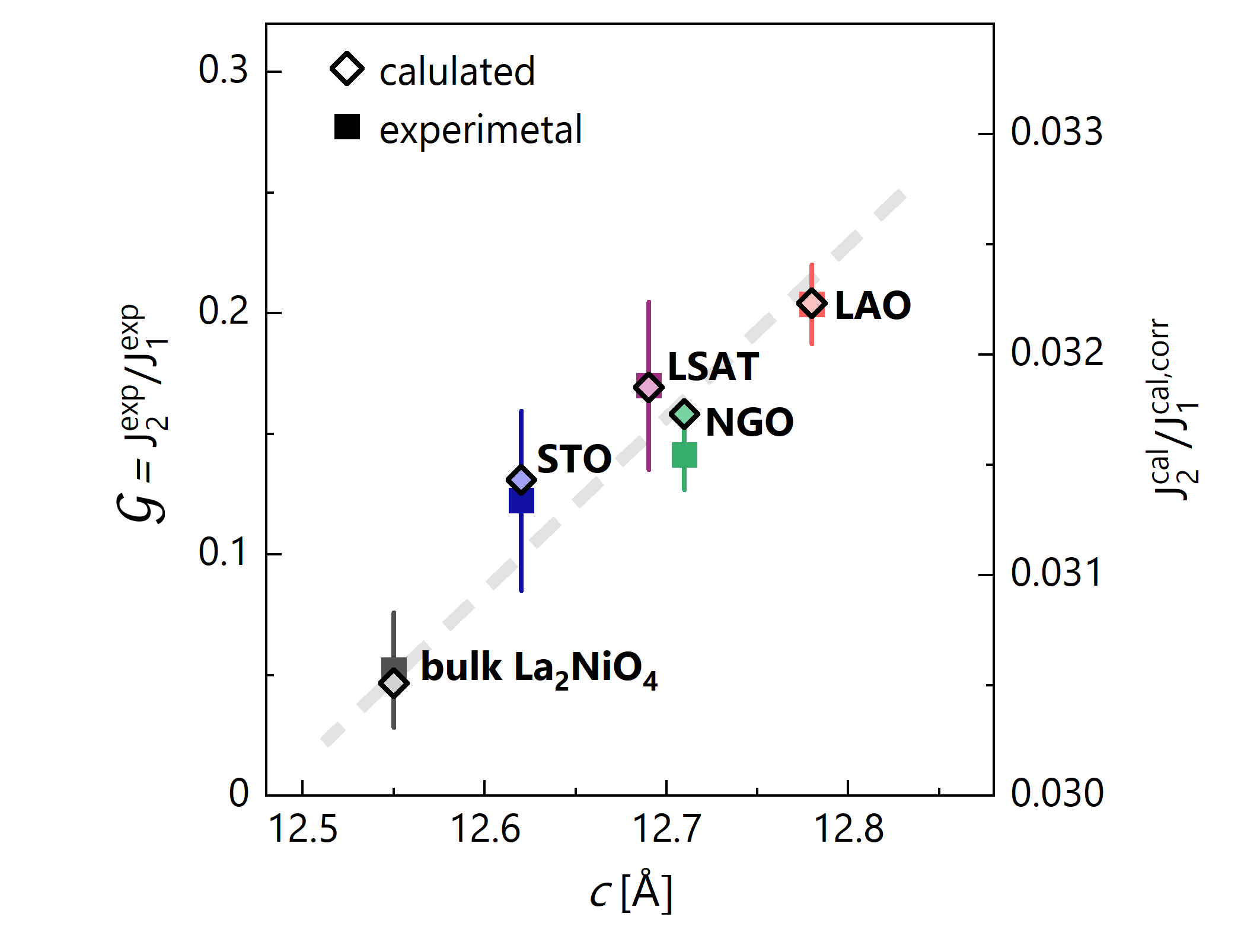}
\caption{\textbf{Strain dependent magnetic frustration.} 
The frustration parameter $\mathcal{G}=J_2^{\mathrm{exp}}/J_1^{\mathrm{exp}}$ derived from the experimental data (squares; left axis) is presented as a function of the \textit{c}-axis lattice parameter. The error bars for the experimental data are calculated as a propagation of standard deviations extracted from the fits. The results for films are combined with data for bulk La$_2$NiO$_4$ from Ref.~\onlinecite{Petsch2023}. The experimental frustration $\mathcal{G}$ is compared to the ratio $J_2^{\mathrm{cal}}/J_1^{\mathrm{cal,{corr}}}$ derived from the DFT and cRPA calculations (diamonds; right axis). Note that the calculated $J_2^{\mathrm{cal}}$ only contains a contribution to the full next nearest-neighbor coupling $J_2$. Therefore, the comparison merely highlights a similar trend of the frustration under in-plane compression. The dashed line is a guide-to-the-eye.}
\label{Fig4}
\end{figure}

As a first step, we parameterize the magnon dispersion of \LNO using a phenomenological spin-wave model that includes effective NN and NNN exchange interactions, respectively $J_1$ and $J_2$~(Fig.~\ref{fig_SpinWave}), plus an easy-plane anisotropy $K$, already reported by previous measurements \cite{nakajima1993spin, Petsch2023}. As a starting point, we employ the AF structure of the bulk \LNO\ determined by neutron diffraction \cite{aeppli1988magnetic, yamada1992magnetic, rodriguez1991neutron}, with the spin direction parallel to the crystallographic $a$-axis. The model is solved in a linear spin-wave (large-$S$) limit, and the calculated dispersion is fitted to the measured one (see Methods). Fitting the experimental (exp) data yields an effective NN exchange interaction $J_1^{\mathrm{exp}}\sim 30$~meV consistent with previous neutron and RIXS results~\cite{FabbrisPRL2017,YamadaJPSJ1991}. Due to the demonstrated finite zone-boundary dispersion, our spin-wave model fitting also yields a moderate NNN exchange interaction $J_2^{\mathrm{exp}}$. Importantly, $J_2^{\mathrm{exp}}$ is positive and enhanced by compressive strain~\cite{ColdeaPRL2001,HeadingsPRL2010}. In what follows, we wish to extract the frustration parameter $\mathcal{G}=J_2^{\mathrm{exp}}/J_1^{\mathrm{exp}}$ with the highest precision. Within our spin-wave model, $E_{X}=4SZ_c(J_1-2J_{2})$ and $E_{\Sigma}=4SZ_c(J_1-J_2)$, where $Z_\text{c}$ is the quantum renormalization factor for spin-wave energies, which is taken as $Z_\text{c}=1.09$ \cite{igarashi1992expansion}. This gives $\mathcal{G}^{-1}=1+\nicefrac{E_{\Sigma}}{(E_{\Sigma}-E_{X})}$. The frustration parameter $\mathcal{G}$ is thus derived directly from the experimental data, with high precision ($E_{\Sigma}$ and $E_{X}$ are extracted with error lower than $5$~meV) and plotted as a function of the \textit{c} lattice parameter in Fig.~\ref{Fig4} (see also Table~S1, Supplementary Note~2). Due to the Poisson effect, the \textit{c} lattice parameter undergoes a proportional shrinkage when the in-plane parameters expand. Our x-ray diffraction measurements confirm this relationship (Fig.~\ref{Fig1}b), indicating that the \textit{c}-axis lattice parameter can serve as an indirect probe of the in-plane strain. Therefore, our findings demonstrate a nearly linear correlation between magnetic frustrations and epitaxial strain.

\begin{table*}[]
  \begin{tabular}{c|cc|rrrr|rr|rrrr|rrr}   
  \hline \hline
    System &  $a$
    &  $c$
    &  \hspace{3mm} $\Delta_{eg}$     
    &  \hspace{3mm} $t_{\alpha\alpha}$   
    &  \hspace{3mm}  $t_{\beta\beta}$    
    &  \hspace{3mm} $t_{\alpha\beta}$    
    &  \hspace{3mm} $t_{\alpha\alpha}^\prime$  
    &  \hspace{3mm} $t_{\beta\beta}^\prime$  
    &  \hspace{3mm}    $U_{\alpha\alpha}$ 
    &  \hspace{3mm} $U_{\beta\beta}$  
    & \hspace{3mm}  $U_{\alpha\beta}$   
    &  \hspace{3mm}$J_H$
    &  $J_1^{\mathrm{cal}}$
    &  $J_2^{\mathrm{cal}}$ 
    &  $J_1^{\mathrm{cal,corr}}$
    \\ &[\AA]&[\AA]&[eV]&[eV]&[eV]&[eV]&[meV]&[meV]&[eV]&[eV]&[eV]&[eV]&[meV]&[meV]&[meV]\\
    \hline
    \LNO&3.890&12.55&0.48&-0.070&-0.403&-0.161&-8.6&74.9&3.06&3.15&1.97&0.52&60.7&1.57&51.0\\
    \hline
    LNO/STO&3.905&12.62&0.48&-0.067&-0.395&-0.156&-8.3&74.5&3.01&3.14&1.94&0.52&59.5&1.56&49.7\\
    LNO/LSAT&3.868&12.69&0.55&-0.065&-0.410&-0.156&-7.4&76.3&3.00&3.11&1.92&0.51&62.2&1.65&51.8\\
    LNO/NGO&3.859&12.71&0.58&-0.064&-0.414&-0.155&-7.1&76.7&3.03&3.16&1.97&0.51&62.3&1.65&51.9\\
    LNO/LAO&3.793&12.78&0.74&-0.060&-0.447&-0.156&-5.5&80.5&3.02&3.08&1.93&0.50&71.6&1.83&56.8\\   
    \hline \hline
\end{tabular}  
\caption{ \textbf{Parameters of the two-orbital Hubbard model.} Crystal field splitting $\Delta_{eg}$, (next-)nearest neighbor hopping $t_{ij}^{(\prime)}$ between $i$-th and $j$-th Ni orbitals ($\alpha$ and $\beta$ here denote the $z^2$ and $x^2-y^2$ orbital, respectively), the  inter- and intra-orbital Coulomb interaction $U_{ij}$, and Hund's exchange $J_H$ between the two orbital as calculated  by  DFT and cRPA  with the in-plane lattice constant $a$ of the three substrates; note that $t_{\alpha\beta}^\prime=0$ by symmetry. From these \textit{ab initio} calculated parameters the spin couplings $J_1$ and $J_2$ are calculated from super-exchange (second order perturbation theory), i.e., from Eq.~(\ref{eq:J}) with $t$ and $t'$, respectively. Estimating higher order terms using a one-orbital analogy, yields the reduced $J_1^{\mathrm{cal,corr}}$ couplings -- see text. The bulk lattice parameters refer to the low temperature tetragonal polymorph of La$_2$NiO$_4$, after Ref.~\onlinecite{Petsch2023}.}
\label{table:theo}
\end{table*}

We stress that, for interaction strengths and hoppings that are realistic for cuprates and nickelates, $J_2>0$ is hard to reconcile with a single-band Hubbard model. A positive $J_2$ implies an effective AF NNN exchange interaction, at odds with what is observed in cuprates~\cite{ColdeaPRL2001, peng2017influence} and $d^9$ infinite-layer nickelates~\cite{LuSci2021,Gao22}. Both systems have indeed been successfully described using a single $d$-orbital framework~\cite{ColdeaPRL2001,HeadingsPRL2010,DelannoyPRB2009,PiazzaPRB2012}. Therefore, we argue that the magnon zone boundary dispersion in LNO signals physics beyond the single-orbital Hubbard model. We propose that the multi-orbital (\dx, \dz) nature of nickelates~\cite{HorioNC2018,UchidaPRB2011,UchidaPRL2011} must be explicitly considered. Already in \LCO, due to the short apical oxygen distance, a small but significant orbital hybridization between \dz\space and \dx\space has been reported~\cite{MattNC2018}. In \LNO\ the apical oxygen distance is even shorter, as exemplified by the reduced $c$ lattice parameter (see Fig.~\ref{Fig1}b), and hence an even more pronounced hybridization is expected.   

To rationalize the trend in the exchange interactions obtained from our spin-wave fits, we derive a two-orbital low-energy model for La$_2$NiO$_4$ on different substrates from first principles (see the Method section). For the Ni $d_{z^2}$ and $d_{x^2-y^2}$ orbitals (labeled $\alpha$ and $\beta$), we compute the (next) nearest-neighbor hopping parameters $t^{(\prime)}$, the crystal field splitting $\Delta_{eg}$, local Coulomb (Hubbard) interaction $U$ and Hund's exchange $J_H$ using experimental lattice constants from Table~\ref{table:theo}. Noteworthy\cite{cRPA_pressure}, the hopping parameters and Coulomb interactions, listed in Table~\ref{table:theo}, hardly change under varying in-plane compression. This is different from calculations for the cuprate family, see Ref.~\onlinecite{IvashkoNC2019}, and agrees with our experiments, which show substantially smaller changes in the magnon spectrum than for the cuprates. What is most affected by strain in Table~\ref{table:theo} is the crystal-field splitting $\Delta_{eg}$ by which the Ni $d_{x^2-y^2}$ orbital is higher in energy than the $d_{z^2}$ orbital. When going from the STO to the LAO substrate, in-plane strain pushes the $d_{x^2-y^2}$ orbital further up in energy, as it is pointing towards the now closer in-plane oxygen sites that are charged negatively.

This crystal-field splitting $\Delta_{eg}$ enters the calculated (cal) two-orbital superexchange as follows:
\begin{equation}
   J_1^{\mathrm{cal}} = \frac{t_{\alpha\beta}^2}{U+J_H-\Delta_{eg}} + \frac{t_{\alpha\beta}^2}{U+J_H+\Delta_{eg}}+ \frac{t_{\alpha\alpha}^2+t_{\beta\beta}^2}{U+J_H} \; ,
   \label{eq:J}
 \end{equation}
where we extend the formula of Ref.~\onlinecite{Lemanski18} to finite $\Delta_{eg}$ (see Supplementary Note~3). As $\Delta_{eg}$ appears once with a plus and once with a minus sign in the denominator, the  crystal-field splitting enters $J_1^{\mathrm{cal}}$ in a higher-than-linear order. 

The magnetic exchange couplings $J_1^{\mathrm{cal}}$ determined by Eq.~(\ref{eq:J}) are displayed in Table~\ref{table:theo}. They show the same qualitative tendency as in our experiment, i.e., an increase of both $J_1^{\mathrm{exp}}$ (see Fig.~\ref{fig_SpinWave}) and $J_1^{\mathrm{cal}}$ with compressive strain. Quantitatively, the {\it ab initio} calculated $J_1^{\mathrm{cal}}$ is however too large. This has two major origins: (i) The cRPA interactions are, here, taken at zero frequency $U=U(\omega=0)$. Additional renormalizations from the frequency dependence $U(\omega)$ \cite{PhysRevB.70.195104} are often mimicked through an empirical enhancement of $U$. Increasing $U$, we could easily obtain quantitative agreement with the experimental $J_1$, but at the cost of a free fit parameter and most likely only an accidental agreement. (ii) Eq.~(\ref{eq:J}) only includes terms to second-order perturbation theory in $t$. For the one-band Hubbard model, higher-order processes have been calculated and yield a correction from $J_1= 4t^2/U$ to a reduced $J_1 = 4(t^2/U- 16 t^4/U^3)$~\cite{MacDonald90,DelannoyPRB2009}. Higher-order terms are expected to reduce $J_1$ also in the two-orbital setting. Here, we estimate these corrections by the one-orbital prescription $t^2/U\longrightarrow t^2/U- 16 t^4/U^3$. Then, e.g., for the NGO substrate, the leading contribution to $J_1^{\mathrm{cal}}$, i.e., $t_{\beta\beta}^2/(U_{\mathrm{eff}})$ with $U_{\mathrm{eff}}=U+J_H$, reduces from 47$\,$meV to 37$\,$meV. Applying this substitution to Eq.~(\ref{eq:J}) yields the corrected exchange couplings $J_1^{\mathrm{cal,corr}}$ listed in Tab.~\ref{table:theo}, which are in better agreement with the measured values (see Fig.~\ref{fig_SpinWave}).

Table~\ref{table:theo} also lists the next-nearest neighbor exchange $J_2^{\mathrm{cal}}$ 
that can be obtained with the same second-order formula  Eq.~(\ref{eq:J}), except now using the hoppings $t^\prime$ instead of $t$ (see Supplementary Note~3 for details). Crucially, our calculations predict a positive $J_2$, in agreement with the experiment. Moreover, $J_2^{\mathrm{cal}}$  shows the same qualitative tendency as in the experiment as a function of  epitaxial strain. On a quantitative level, the calculated values are, however, a factor $\sim3-5$ lower than the experimental results. The reason is that contributions to $J_2$ from higher-order exchange processes, of order $t^4/U^3$ and ${t^\prime}t^2/U^2$, become (relatively) more important for $J_2^{\mathrm{cal}}$, as the second-order terms are now based on the much smaller ${t^\prime}$. 

The key difference between the multi-orbital case of LNO and the one-orbital cuprates is the larger $U$ ($U_{\mathrm cRPA}\approx 3.1\,$eV for two-orbitals while $U_{\mathrm cRPA}\approx 1.9\,$eV for a one Ni $d_{x^2-y^2}$  orbital setup) and the additional Hund's $J_H\approx0.5\,$eV in the denominator of Eq.~(\ref{eq:J}). As a consequence, the balance for the NNN exchange coupling shifts from a  ferromagnetic ring exchange $J_{2} \sim - t^4/U_{\mathrm eff}^3<0$, that overpowers the AF second-order exchange $J_2 \sim  {t^\prime}^2/U_{\mathrm eff}>0$  in the one-orbital cuprates, toward dominance of the latter in the multi-orbital LNO. This change in hierarchy explains the main qualitative differences in magnon dispersion between LNO and cuprates: the opposite sign of the effective $J_2$. For LNO, with a positive $J_2$, the zone-boundary dispersion shows a notable minimum at $(\nicefrac{1}{2},0)$, see Fig.~\ref{fig_SpinWave}, whereas a maximum occurs for the negative $J_2$ in cuprates.\\

\noindent\textbf{Conclusions} \\
The \textit{ab-initio} calculations indicate that the magnetic frustration in La$_2$NiO$_4$ is caused by the multi-orbital nature of 3$d^8$ nickelates. More importantly, our results demonstrate that the degree of frustration is amplified by compressive strain (see Fig.~\ref{Fig4}), with a pivotal role played by the crystal-field splitting. Indeed, with the substrates used, the magnetic frustration increases four-fold with respect to the bulk, bridging half the way toward the exotic realm anticipated for $J_2/J_1\sim \nicefrac{1}{2}$. 
Thus, our study suggests an effective tool for tuning antiferromagnetic interactions within square lattice systems. We speculate that the approach is applicable beyond La$_2$NiO$_4$ and
may offer an experimental route to reach so far unexplored regions of the magnetic phase diagram, potentially allowing  to investigate exotic states induced by magnetic frustration. 
\\[2mm]

\noindent\textbf{Methods} \\
\noindent\textbf{Film growth and characterization.} Thin films of La$_2$NiO$_4$ were grown by RHEED-equipped Radio-frequency off-axis magnetron sputtering~\cite{Podkaminer2016} on (001) STO, (001) LAO, (001) LSAT and (110) NGO substrates. These films were grown in an argon atmosphere at $700^{\circ}$C. Their qualities were confirmed by atomic force microscopy and x-ray diffraction. Their insulating character was confirmed by resistivity measurements of the LNO/STO film (see Fig.~S3, Supplementary Note~4).\\

\noindent\textbf{RIXS experiments.} Ni $L$-edge RIXS experiments for STO, LAO and NGO substrates were carried out at the I21 beamline~\cite{Zhou2022} at the DIAMOND Light Source. All spectra were collected in the grazing exit geometry using linear horizontal polarized incident light with the scattering angle fixed to $2\theta=154^{\circ}$. The energy resolution was estimated from the elastic scattering on amorphous carbon tape and was between $37$-$41$~meV (full-width-at-half-maximum, FWHM). All films were measured at base temperature $T=16$~K. We define the reciprocal space $(q_x,q_y,q_z)$ in reciprocal lattice units $(h,k,\ell)=(q_{x}a/2\pi,q_{y}b/2\pi,q_{z}c/2\pi)$ where $a,b$ and $c$ are the pseudo-tetragonal lattice parameters. RIXS spectra were acquired along three in-plane paths: $(0,0)\rightarrow(0,\nicefrac{1}{2})$, $(0,0)\rightarrow(\nicefrac{1}{4},\nicefrac{1}{4})$ and $(0,\nicefrac{1}{2})\rightarrow(\nicefrac{1}{4},\nicefrac{1}{4})$. Extraction of low-energy excitations around $(0,0)$ is limited by energy resolution. Due to kinematic constraints $\Gamma$ points at higher zones cannot be reached, as well. RIXS intensities are normalized to the weight of the $dd$ excitations~\cite{WangPRL2020}. The data for the LSAT substrate were collected at the ID32 beamline at the European Synchrotron Radiation Facility (ESRF) (see description in Supplementary Note~5, Fig.~S4). \\ 

\noindent\textbf{Phenomenological spin-wave model.}
The effective superexchange parameters were extracted from the measured dispersion using a linear spin wave model. We included effective couplings between the first and second nearest neighbours, plus an easy-plane anisotropy $K$, with the resulting Hamiltonian:
\begin{equation}
    \mathcal{H} = J_{1}\sum_{\langle i,j\rangle} \mathbf{S}_i \cdot \mathbf{S}_j + J_{2}\sum_{\langle\langle i,j\rangle\rangle} \mathbf{S}_i \cdot \mathbf{S}_j + K\sum_i (S^z_i)^2
\end{equation}
where $\langle i,j \rangle$ and $\langle\langle i,j \rangle\rangle$ denote pairs of first and second nearest neighbours Ni atoms, respectively. The fitting procedure has been carried out using the SpinW package~\cite{toth2015spinW}. As an input we have used the AF structure of the bulk \LNO\ determined by neutron diffraction~\cite{aeppli1988magnetic, yamada1992magnetic, rodriguez1991neutron}, with the spin direction parallel to the crystallographic $a$-axis. The dispersion in the approximation of the linear spin wave theory is represented by~\cite{ColdeaPRL2001, Petsch2023}:
\begin{equation}
  \begin{split}
    \hbar \omega & = Z_c \sqrt{(A_\textbf{q}^2 - B_\textbf{q}^2)} \\
    A_\textbf{q} & = 4S \biggl [\frac{K}{4} + J_1 - J_2(1-\nu_h \nu_k) \biggr ] \\
    B_\textbf{q} & = 4S \biggl [ J_1 \, \frac{\nu_h + \nu_k}{2} - \frac{K}{4} \biggr ] 
    \end{split}  
\end{equation}
where $\nu_x = \cos(2\pi x)$. The quantum renormalization factor for spin-wave velocity is fixed to $Z_c=1.09$, as usual for $S=1$ systems~\cite{igarashi1992expansion}. The value of the easy-plane anisotropy $K$ mostly controls the size of the magnon gap at the $\Gamma$ point. Since this value is very hard to obtain from RIXS spectra, we have fixed $K=0.5$~meV in agreement with previous inelastic neutron scattering data \cite{nakajima1993spin}. We have also neglected other interactions $<10^{-1}\,$meV such as easy-axis anisotropy, inter-layer coupling and Dzyaloshinskii–Moriya interactions~\cite{Petsch2023, yamada1992magnetic}.\\

\noindent\textbf{\textit{Ab initio} calculations.} Electronic structure calculations were performed with density functional theory in the local density approximation using a full-potential linearized muffin-tin orbital (FPLMTO) code~\cite{fplmto}, after the structures were optimized with WIEN2k~\cite{wien2k2020} using the PBE functional. 
We mimicked the influence of the substrates by simulating bulk La$_2$NiO$_4$ using the experimental lattice constants of the thin films. The reference calculation for the bulk uses lattice constants from Ref.~\onlinecite{Petsch2023}. All calculations assume the space group I4/mmm and are paramagnetic. The resulting band-structures are displayed in Fig.~S5, Supplementary Note~6. The FPLMTO calculations were converged using $12^3$ reducible $k$-points and include local orbitals for the Ni-3p and La-5p states. The internal atomic positions were relaxed with WIEN2k using $6^3$ reducible $k$-points, a cutoff parameter $\hbox{RMTKMAX}=7$ and partial waves inside the atomic spheres up to $l=5$, until the forces were below 1mRy per Bohr radius (for details of the relaxed structures, see Supplementary Note~7, Table~S2). The tight-binding hopping and crystal-field parameters have been extracted from a projection onto maximally localized Wannier orbitals~\cite{RevModPhys.84.1419,miyake:085122} of Ni $3d_{x^2-y^2}$ and $3d_{z^2}$ character. Matrix elements of the static ($\omega=0$) and local screened Coulomb interaction (Hubbard $U$ and Hund's $J_H$) have been estimated from calculations in the constrained random phase approximation (cRPA)~\cite{PhysRevB.70.195104} for entangled band-structures~\cite{miyake:155134} in the Wannier basis~\cite{miyake:085122} using $6\times6\times6$ reducible momentum-points in the Brillouin zone. For the two-particle product basis, states are kept up to an angular cutoff of $l=4$ and down to an overlap eigenvalue of $10^{-4}$. \\[2mm]

\noindent\textbf{Data availability}\\ 
Data supporting the findings of this study are available from corresponding authors upon reasonable request. \\[2mm]

\noindent\textbf{Code availability}\\ 
Code supporting the data processing of this study is available from corresponding authors upon reasonable request. \\[2mm]

\noindent\textbf{Author contributions}\\
I.B. and L.M. have contributed equally to this work. G.D.L., C.B.E., A.D., L.G., A.E., S.J. and M.G. grew and characterized the \LNO\ films. I.B., J.Choi, M.G-F., S.A., K-J.Z, K.K., N.B.B. and Q.W. carried out the RIXS experiments. I.B., L.M. and Q.W. analysed the RIXS data. P.W., J.M.T and K.H. conceived, executed and analysed the \textit{ab initio} calculations. I.B., Q.W. and J.C. conceived the project. All authors contributed to the writing of the manuscript.\\[2mm]

\noindent\textbf{Acknowledgements}\\
I.B. and L.M. acknowledge support from the Swiss Government Excellence Scholarship under project numbers ESKAS-Nr:~2022.0001 and ESKAS-Nr:~2023.0052. G.D.L, M.G. and J.C. thank the Swiss National Science Foundation under Projects No.~200021$\textunderscore$188564 and PP00P2$\textunderscore$170564. Q.W. is supported by the Research Grants Council of Hong Kong (ECS No.~24306223), and the CUHK Direct Grant (No.~4053613). G.D.L. acknowledges support from the UZH GRC Travel Grant. We acknowledge Diamond Light Source for providing beam time on beamline I21 under Proposal MM30189 and the European Synchrotron Radiation Facility (ESRF) for providing beam time on beamline ID32 under Proposal HC~5241. C.B.E. acknowledges support for this research through a Vannevar Bush Faculty Fellowship (ONR N00014-20-1-2844), the Gordon and Betty Moore Foundation’s EPiQS Initiative, Grant GBMF9065 and NSF through the University of Wisconsin Materials Research Science and Engineering Center (DMR-2309000). The synthesis of thin films at the University of Wisconsin-Madison was supported by the US Department of Energy (DOE), Office of Science, Office of Basic Energy Sciences (BES), under award number DE-FG02-06ER46327.   \\[2mm]

\noindent\textbf{Competing interests} \\
The authors declare that they have no competing interests.

\def\bibsection{\section*{\refname}} 
\bibliography{lsco_ref}

\includepdf[pages={{},1,{},2,{},3,{},4,{},5,{},6}]{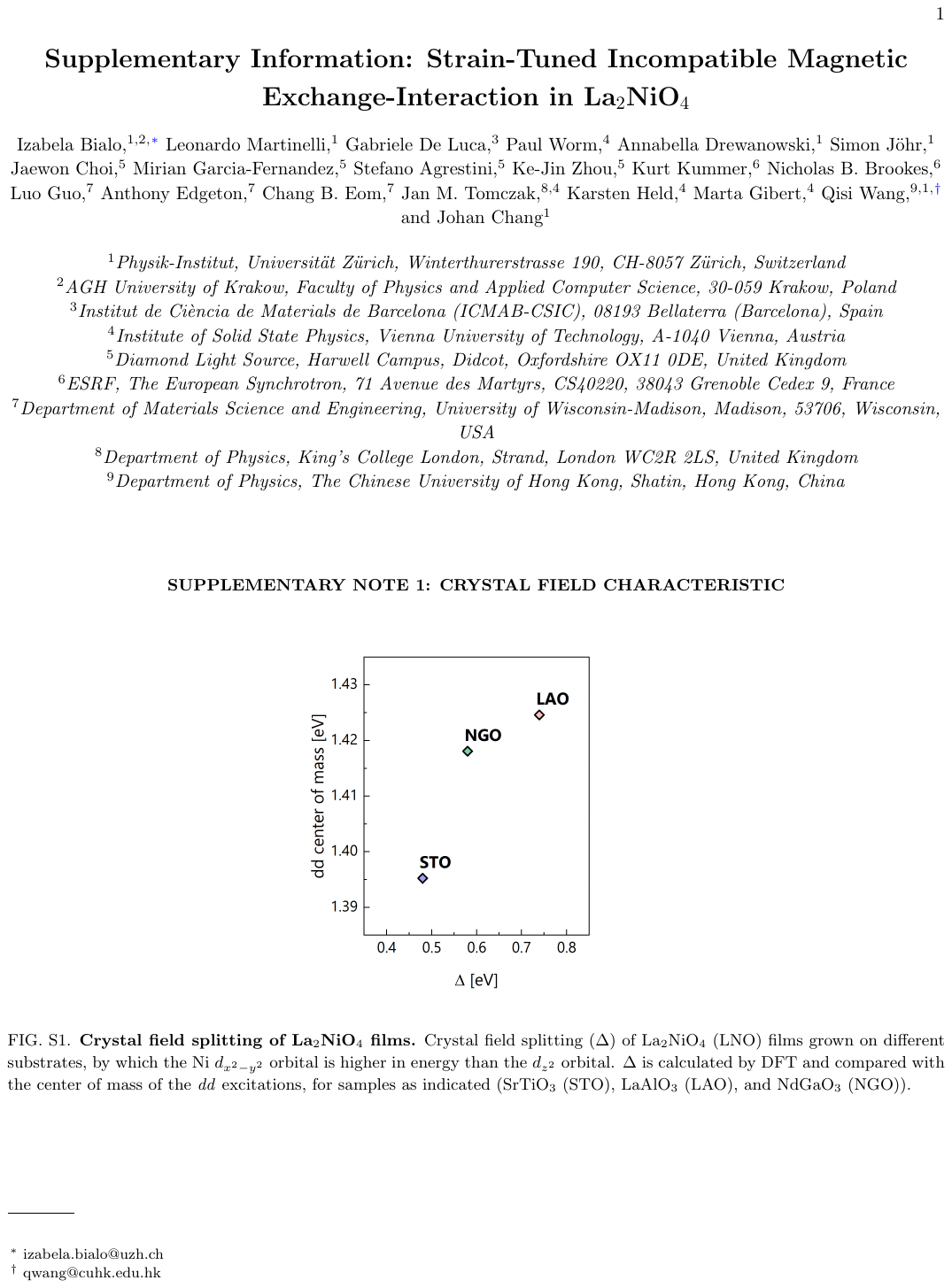}
\end{document}